\documentclass[prd,aps,nofootinbib,showpacs]{revtex4}
\usepackage{graphicx}
\begin{document}

\title{High velocity spikes in Gowdy spacetimes}

\author{David Garfinkle}
\email[Email:]{garfinkl@oakland.edu}
\affiliation{Department of Physics, Oakland University
Rochester, Michigan 48309}
\author{Marsha Weaver}
\email[Email:]{mweaver@phys.ualberta.ca} 
\affiliation{Theoretical Physics Institute, University of Alberta
Edmonton, AB, Canada T6G 2J1}

\begin{abstract}
We study the behavior of spiky features in Gowdy spacetimes. 
Spikes with velocity initially high are, generally, driven to
low velocity.
Let $n$ be any integer greater than or equal to 1.  If the
initial velocity of an upward pointing spike is between $4n-3$
and $4n-1$ the spike persists with final velocity between 1 and 2,
while if the initial velocity is between $4n-1$ and $4n+1$, the
spiky feature eventually disappears.  For downward pointing
spikes the analogous rule is that spikes with initial velocity
between $4n-4$ and $4n-2$ persist with final velocity between
0 and 1, while spikes with initial velocity between 
$4n-2$ and $4n$ eventually disappear.
\end{abstract}
\pacs{04.20.-q, 04.25.Dm, 04.20.Dw}
\maketitle
\section{Introduction}

There have been several investigations of the approach to the singularity
in inhomogeneous cosmologies, predominantly in the case that a two-dimensional
symmetry group acts spatially \cite{vinceandjim,piotr1,piotr2,boroandvince,
beverlyandvince,beverlyandme,mjb,alan,alanandmarsha,bjm,wainwright,ringstrom}.
The most extensively studied class of 
such spacetimes is the class of Gowdy spacetimes\cite{gowdy} 
on ${T^3} \times R$.
Numerical studies of these spacetimes show that the approach to the singularity
is asymptotically velocity term dominated (AVTD) except at an isolated
set of points.  

As shown in \cite{beverlyandme} this behavior can be understood by
considering certain terms in the equations as ``potentials'' that 
affect the dynamics at each spatial point.  These potentials drive
the dynamics into the AVTD regime, except at those isolated points
where one of the potentials vanishes.  The behavior at each of these
isolated points differs from that of its neighbors, leading to the 
creation of features called spikes that become narrow
as the singularity is approached.  The analysis of \cite{beverlyandme}
uses an explicit approximation to find a closed form
expression for the behavior of  the spikes.  However,
this approximation only remains valid throughout the approach
to the singularity for a certain
class of spikes: those for which the ``velocity'' $v$ of the spike
is ``low'' ($1<v<2$ for ``upward pointing spikes'' and $0<v<1$
for ``downward pointing spikes'').  What then happens if a spike forms
and it initially  has ``high'' velocity?  Certain terms in the evolution
equations which are decaying for low velocity spikes are instead growing
in magnitude for a high velocity spike and the sign of the net effect
is such that the velocity should decrease \cite{marsha,alanandmarsha}.
An explicit approximation that is valid as the velocity at a high
velocity spike decreases has not been found, leaving open 
the question, ``What is the long term behavior of
initially high velocity spikes in Gowdy spacetimes?''

To answer this question, we perform numerical simulations of
Gowdy spacetimes with these high velocity spikes.  Since the
spikes become ever narrower as the singularity is approached,
until just this narrowness causes the velocity to decrease
if it is originally high,
extremely high resolution is needed to follow the detailed
behavior of the spike.  We use a numerical
method that was used in \cite{dgcrit} to study critical collapse.  
Here the equations are put in characteristic form and the outermost 
grid point is chosen to be the ingoing light ray that hits the singularity
at the center of the spike.  In this way, the grid gets smaller as the
feature that it needs to resolve shrinks.

Section 2 presents the equations and the characteristic numerical
method.  Results are given in section 3 and conclusions in section 4.      

\section{Equations and numerical methods}

The Gowdy metric on ${T^3} \times R$ takes the form
\begin{equation}
d{s^2}={e^{\lambda /2}}{t^{-1/2}}(- d{t^2} + d {x^2}) + t [ {e^P}
{{(dy +Q d z)}^2} + {e^{-P}} d{z^2}]
\end{equation}
where $P,\, Q$ and $\lambda $ are functions of $t$ and $x$.  The vacuum
Einstein field equations split into ``evolution'' equations for
$P$ and $Q$
\begin{eqnarray}
{P_{,tt}} + {t^{-1}} {P_{,t}} - {P_{,xx}} + {e^{2P}}( {Q_{,x} ^2} - 
{Q_{,t} ^2}) = 0 
\label{evolveP}
\\
{Q_{,tt}}+{t^{-1}} {Q_{,t}}- {Q_{,xx}} + 2 ( {P_{,t}}{Q_{,t}} - {P_{,x}}
{Q_{,x}})=0 
\label{evolveQ}
\end{eqnarray}
and ``constraint'' equations for $\lambda$
\begin{eqnarray}
{\lambda _{,t}}=t [ {P_{,t} ^2} +{P_{,x} ^2} +{e^{2P}} ( {Q_{,t} ^2}
+ {Q_{,x} ^2})]
\label{lambdat} \\
{\lambda _{,x}}=2 t ( {P_{,x}}{P_{,t}}+{e^{2P}}{Q_{,x}}{Q_{,t}})
\label{lambdax}
\end{eqnarray}
(here ${_{,a}}=\partial /\partial a$).
The constraint equations determine $\lambda$ once $P$ and $Q$ are known.
The integrability conditions for the constraint equations are satisfied
as a consequence of the evolution equations.  Since the evolution equations
do not depend on $\lambda$ there is essentially a complete decoupling
of constraints from evolution equations.  Therefore, for the purposes of this
paper we will treat only equations (\ref{evolveP}-\ref{evolveQ}).
The only restriction that the constraints place on initial data for
equations (\ref{evolveP}-\ref{evolveQ}) is the following: since
$\lambda $ at $x=0$ is the same as $\lambda $ at $x=2 \pi$, it follows that
the integral from $0$ to $2\pi$ of the right hand side of equation
(\ref{lambdax}) must vanish.  We require that this restriction is satisfied by
the initial data for equations (\ref{evolveP}-\ref{evolveQ}) and then
these equations insure that the restriction is also satisfied at
subsequent times.

The singularity is at $t=0$.  It is often helpful to introduce the
coordinate $\tau \equiv - \ln t$.  Thus the singularity is approached
as $\tau \to \infty$.  In terms of this coordinate, the evolution
equations (\ref{evolveP}-\ref{evolveQ}) become
\begin{eqnarray}
{P_{,\tau \tau}} - {e^{2P}} {Q_{,\tau} ^2} - {e^{- 2 \tau}} {P_{,xx}}
+ {e^{2(P-\tau )}} {Q_{,x} ^2} = 0
\label{tauevolveP}
\\
{Q_{,\tau \tau}} + 2 {P_{,\tau}}{Q_{,\tau}} - {e^{-2\tau}} \left (
{Q_{,xx}}+ 2 {P_{,x}}{Q_{,x}}\right ) = 0
\label{tauevolveQ}
\end{eqnarray} 
The velocity is defined as
$v = \sqrt{P_{,\tau}^2 + e^{2P}Q_{,\tau}^2}$.
In the limit of large $\tau$, if we neglect the terms in equations
(\ref{tauevolveP}-\ref{tauevolveQ}) proportional to $e^{-2 \tau}$ 
then we find that $v \to v_\infty$ and $P \to v_\infty \tau$
where $v_\infty$ depends only on $x$.  However, if
$v_\infty > 1$ then neglect of the term  ${e^{2(P-\tau )}} {Q_{,x} ^2}$
is not justified.  What happens is that if $P_{,\tau}$ is greater than
1, then this term drives it to values less than 1, except at the 
isolated points where ${Q_{,x}}=0$.  At those points upward pointing spikes
form.  The previous studies \cite{beverlyandme,marsha,alanandmarsha} suggest
that if $1 < P_{,\tau} < 2$ at the spike, the spike will persist, but that
if $P_{,\tau} > 2$ at the spike, $P_{,\tau \tau}$ will become negative and
its magnitude will start growing exponentially in $\tau$ as the spike
narrows.  But the previous studies leave open the question, ``What
happens next?''  We want to know the outcome.

For our purposes, we will need the field equations in characteristic form.
To that end we introduce null coordinates $ u \equiv t + x$ and $w \equiv
t - x$ and characteristic variables $A, \, B, \, C$ and $D$ given by
\begin{eqnarray}
A = (u+w) {P_{,w}} 
\\
B = (u+w) {P_{,u}} 
\\
C = (u+w) {e^P} {Q_{,w}} 
\\
D = (u+w) {e^P} {Q_{,u}} 
\end{eqnarray} 
The evolution equations (\ref{evolveP}-\ref{evolveQ}) then become
\begin{eqnarray}
{A_{,u}}= {{(u+w)}^{-1}} [ C D + (A-B)/2 ]
\label{evolveA}
\\
{B_{,w}} = {{(u+w)}^{-1}} [ C D + (B-A)/2 ]
\label{constrainB}
\\
{C_{,u}}= {{(u+w)}^{-1}}  [ - A D + (C-D)/2 ]
\label{evolveC}
\\
{D_{,w}} = {{(u+w)}^{-1}} [ - B C + (D-C)/2 ]
\label{constrainD}
\end{eqnarray}

We impose the condition that $P(t,x)$ and $Q(t,x)$ are even functions
of $x$.  This insures that ${Q_{,x}}(t,0)=0$ and therefore that the 
spike will be at $x=0$.  With this condition, we only need to evolve
on the domain $x \ge 0$.  This corresponds to $u \ge w$.  The 
characteristic initial value formulation for this system is as follows:
on a $u={\rm constant}$ surface we give 
as initial data $A$ and $C$ as functions of
$w$.  Equations (\ref{constrainB}) and (\ref{constrainD}) are then
integrated to yield $B$ and $D$ on the initial data surface.  
The numerical method used is second order Runge-Kutta.  The 
initial condition for this integration is that $B=A$ and $D=C$ at
$u=w$.  This is a consequence of the fact that $P$ and $Q$ are even
functions of $x$.  Now with $A, B, C$ and $D$ known on the initial 
surface, equations (\ref{evolveA}) and (\ref{evolveC}) can be regarded
as ODEs for each value of $w$.  At each grid point, these ODEs are
integrated to yield $A$ and $C$ at the next value of $u$.  The numerical
method used is a second order predictor-corrector method.  This entire
process is iterated to produce the full evolution.

We choose the minimum value of $w$ to be zero, corresponding to the light
ray that will hit the singularity ($t=0$) at the center of the spike
($x=0$).  The maximum value of $w$ is the current value of $u$, 
corresponding to $x=0$.  Thus as the evolution proceeds, some grid points
leave the physical domain and are no longer part of the evolution process.
Throughout, we choose $du=dw$.  This means that we lose one grid point
at each time step.  When half of the grid points are lost, we put them
back in between the remaining ones and interpolate $A$ and $C$ to obtain
their values on the new grid points.  Thus we make the grid twice as fine.
In this way, as the singularity is approached we always have enough grid
points to resolve the ever narrowing spike.     

\section{Results}

While our characteristic method (c-code) allows arbitrary initial data
for $A$ and $C$ (as long as the initial data does not determine a Cauchy
surface\footnote{If a Cauchy surface is determined, then the initial
data must be compatible with the periodicity of $\lambda$.}),
it is helpful to make comparisons with codes using
the usual Cauchy methods.  We therefore also use codes that evolve 
equations (\ref{evolveP}-\ref{evolveQ}) (t-code) and equations 
(\ref{tauevolveP}-\ref{tauevolveQ}) ($\tau$-code).  
These codes use standard centered
differences for spatial derivatives and the iterated Crank-Nicholson
method \cite{matt} (ICN) for time evolution.  We begin with Cauchy
data at $t=\pi$ and evolve using the t-code to
$t=\pi /2$.  In the course of this evolution, we obtain $A$ and $C$
on the null surface given by $u=\pi /2$ and $0 \le w \le \pi /2$, which
we use as initial data for the c-code.    

\begin{figure}
\includegraphics{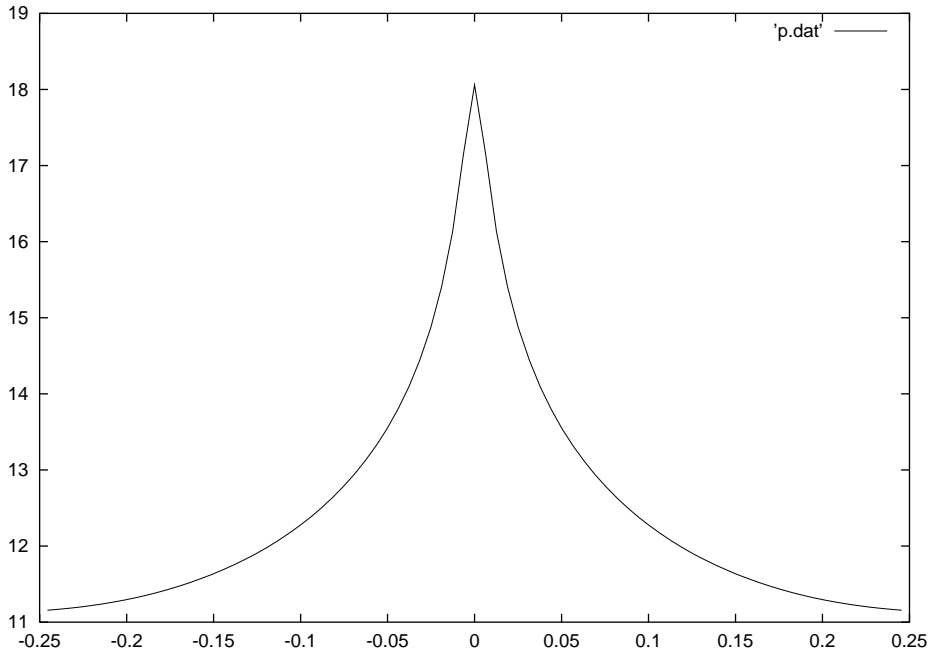}
\caption{\label{f1}spike in $P$ at $\tau =9$}
\end{figure}

\begin{figure}
\includegraphics{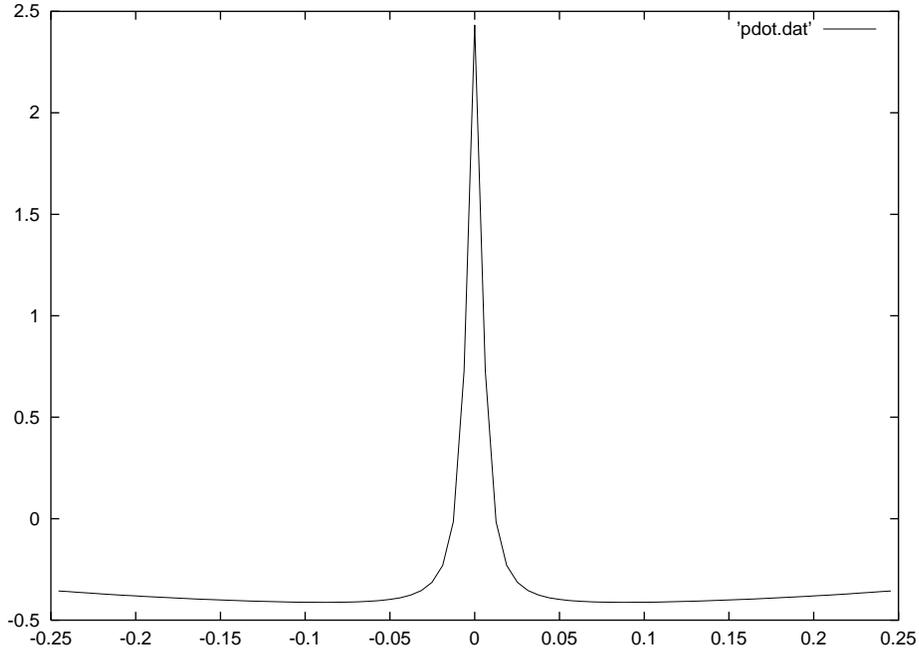}
\caption{\label{f2}spike in $P_{,\tau}$ at $\tau =9$}
\end{figure}

To obtain solutions with high velocity spikes, we note that in the
polarized case ($Q=0$) the velocity can take on any values.  We therefore
choose initial data with small $Q$.  The form of the data is 
$P=0, \; {P_{,\tau}}=s\cos x, \; Q=q\cos x$ and ${Q_{,\tau}}=0$
where $s$ and $q$ are constants.  Figures \ref{f1} and 
\ref{f2} show the results of
evolving these data.  Here $s=-8, \; q=1.0 \times {{10}^{-3}}$
and the result is produced by using the $\tau$-code and evolving to 
$\tau =9$.

\begin{figure}
\includegraphics{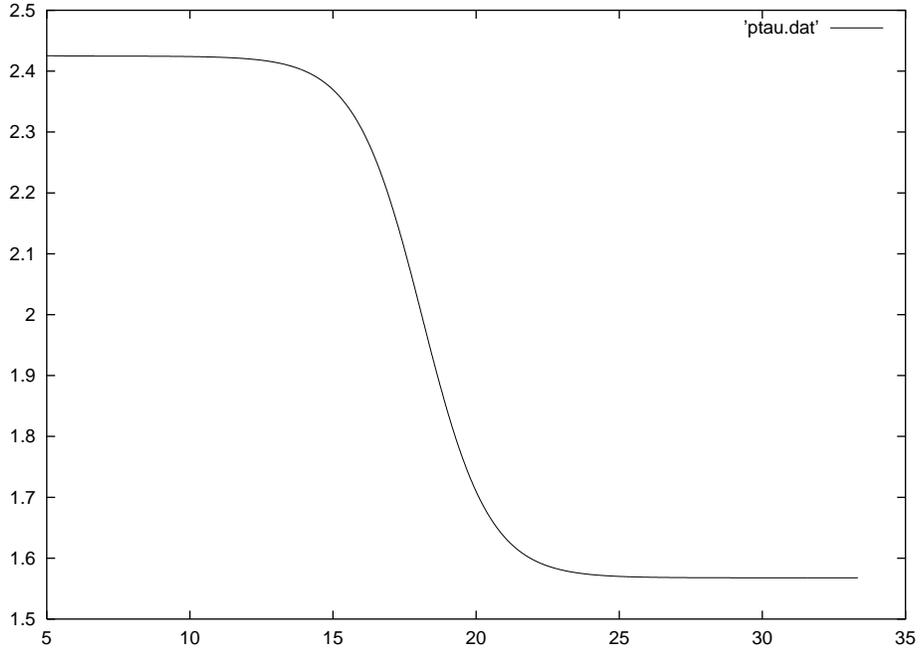}
\caption{\label{f3}$P_{,\tau}$ at the spike}
\end{figure}

Now we take the same initial data and evolve it for a longer time, first
using the t-code to generate the appropriate initial data for the 
c-code and then using the c-code to evolve very close to the singularity.
The result for $P_{,\tau}$ {\it vs} $\tau$ at $x=0$ is
shown in figure \ref{f3}.  As the spike forms, $P_{,\tau} >2$.  As the
spike narrows, $P_{,\tau}$ decreases, as predicted.  Note that
$P_{,\tau}$ at the spike starts out larger than 2, and that it is
eventually driven below 2.

One might wonder what mechanism is responsible for driving the spike
velocity below 2.  The answer comes from an examination of 
equation (\ref{tauevolveP}).  Rewriting this equation with
$P_{,\tau \tau}$ on one side we have 
\begin{equation}
{P_{,\tau \tau}} = {e^{- 2 \tau}} {P_{,xx}} + {e^{2P}} {Q_{,\tau} ^2} 
- {e^{2(P-\tau )}} {Q_{,x} ^2}
\label{vartauevolveP}
\end{equation}
At the spike the term $ {e^{2(P-\tau )}} {Q_{,x} ^2}$ 
vanishes and the term $ {e^{2P}} {Q_{,\tau} ^2}$ 
is positive definite.  The term ${e^{- 2 \tau}} {P_{,xx}}$
is negative, since the spike in $P$ is upward pointing.  
What is happening is that the spike is becoming so narrow and
thus $P_{,xx}$ is becoming so large that despite the smallness of
$e^{- 2 \tau}$ the quantity ${e^{- 2 \tau}} {P_{,xx}}$ is not
negligible.  The term $ {e^{2P}} {Q_{,\tau} ^2}$ is also
not negligible, but its magnitude is less than the magnitude
of ${e^{- 2 \tau}} {P_{,xx}}$, so $P_{,\tau \tau}$ is negative.
In the explicit approximation, $ {e^{2P}} {Q_{,\tau} ^2}$
and ${e^{- 2 \tau}} {P_{,xx}}$ are decaying in magnitude if
$P_{,\tau} < 2$, but if $P_{,\tau} > 2$ and if $Q_{,xx}$ does not
vanish at the spike they are growing in magnitude, and the leading
part of $ {e^{2P}} {Q_{,\tau} ^2}$ is half the magnitude of the
leading part of ${e^{- 2 \tau}} {P_{,xx}}$, so the net effect
is a decrease in $P_{,\tau \tau}$ \cite{beverlyandme,marsha,alanandmarsha}.
From figure \ref{f3} we see that this decrease eventually comes to
an end while $P_{,\tau}$ is still greater than 1.  Now the conditions
for the explicit approximation of the spike to be valid throughout
the approach to the singularity are satisfied, indicating that the
spike will persist.

\begin{figure}
\includegraphics{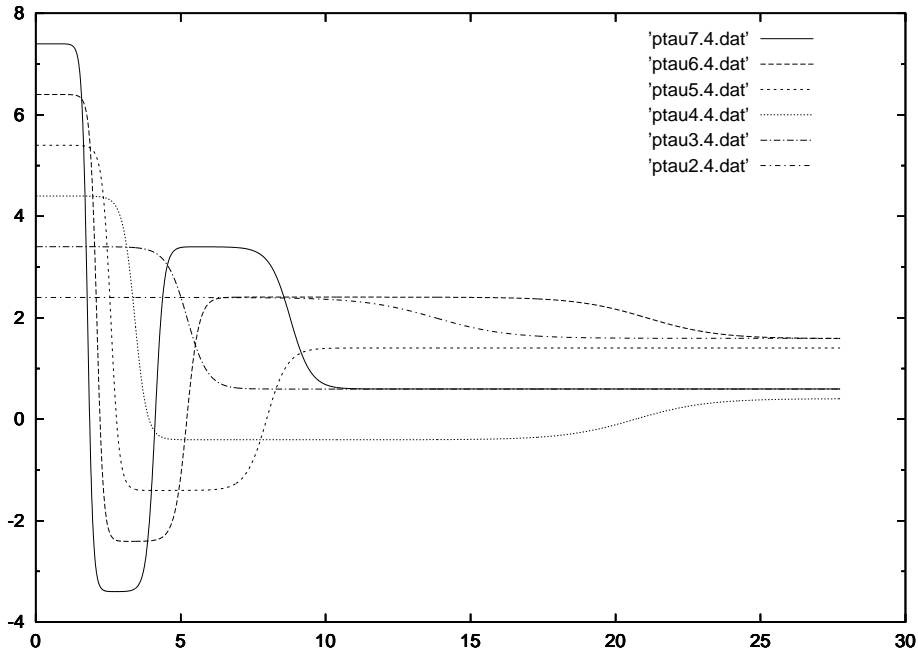}
\caption{\label{figrule}$P_{,\tau}$ at the spike for different initial
values of $P_{,\tau}$.}
\end{figure}

\begin{figure}
\includegraphics{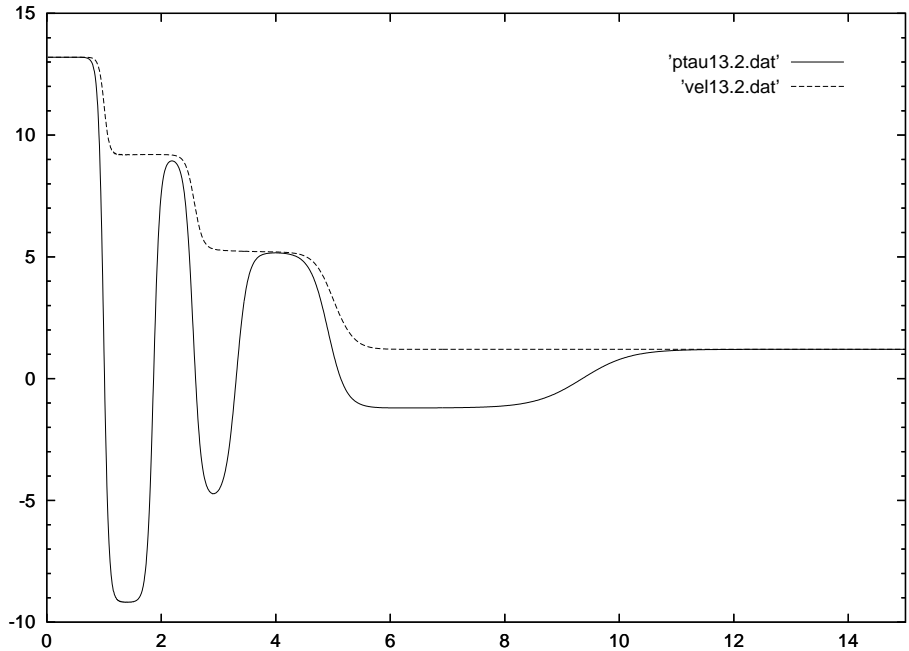}
\caption{\label{figrule2}$P_{,\tau}$ and the velocity
at the spike for $P_{,\tau} =13.2$ initially.}
\end{figure}

\begin{figure}
\includegraphics{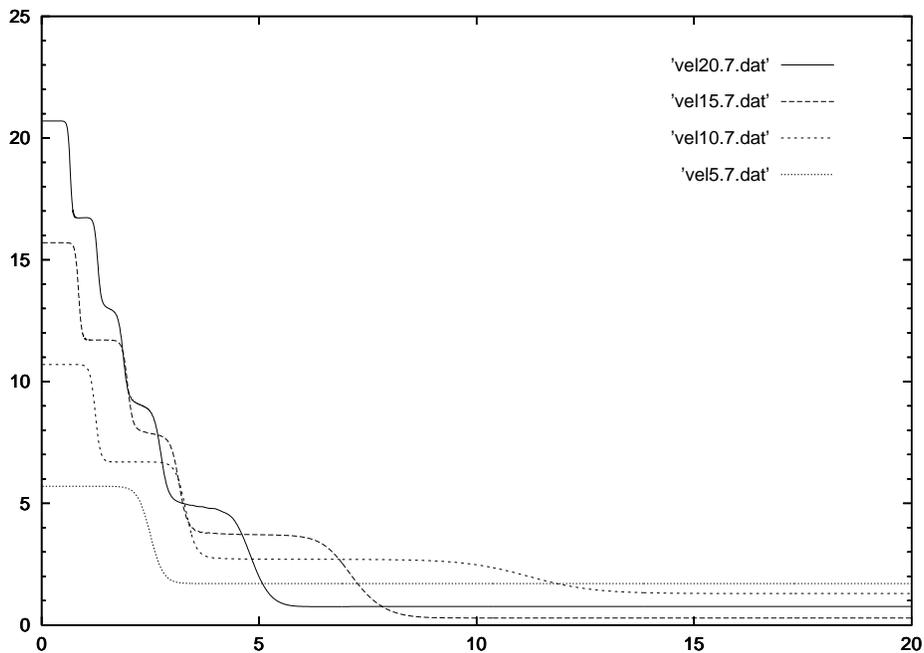}
\caption{\label{figvel}Velocity at the spike for different initial
values $P_{,\tau}$.}
\end{figure}

Another family of initial data which we use to generate spikes is
$t=t_0$, $P=p, \; {P_{,\tau}}=s, \; Q=q\cos x$ and
${Q_{,\tau}}=0$ where $t_0$, $p$, $s$ and $q$ are constants,
with $s >1$.  The constants are chosen so that the explicit
approximation to the spike is initially valid, until the spike
narrows enough that the velocity begins to slide down.
To allow matching of the t-code to the c-code,
the value of $t_0$ is chosen to be $4 \pi n/m$, with $n+1$ the
number of grid points for the c-code and the $m+1$st grid
point identified with the 1st grid point in the t-code (corresponding
to a difference in $x$ of $2 \pi$).  In figures \ref{figrule}-\ref{figvel}
we show results from a family of initial data sets with
$n=160$ and $m=2050$ (which results in $t_0 \approx 0.98$),
$p = 0$ and $q = 0.01$.  We vary $s$, from $2.4$ to $20.7$.
In figure \ref{figrule} we again plot $P_{,\tau}$ {\it vs} $\tau$.
In figure \ref{figrule2} we plot both $P_{,\tau}$ {\it vs} $\tau$
and the velocity, $v$ {\it vs} $\tau$ for a single initial data
set, with $s = 13.2$.
In figure \ref{figvel} we plot $v$ {\it vs} $\tau$.
When $2 < w_0 < 4$ there is just one ``slide'' in the velocity,
$v$.  Its final value is about 4 minus its initial value,
and the final value of $P_{,\tau}$ is about equal to $4 - s$.
When $4 < s < 6$, there is a slide, after which
$P_{,\tau} \approx 4 - s$ ({\it i.e.}, it is negative),
and $v \approx |4 - s|$. After this slide the velocity
is approximately constant, while $P_{,\tau}$ ``bounces'' to a
value equal in magnitude to $4 - s$, but now positive
($|4-s|$).  This sort of bounce has been discussed
in previous work \cite{beverlyandme} and is driven
by the term $ {e^{2P}} {Q_{,\tau} ^2}$ in the evolution equation
for $P$.  When $s > 6$, there
is not just one slide in $P_{,\tau}$, but a number of slides.  Each
slide approximately satisfies the rule $v \to |4-v|$ and $P_{,\tau}
\to 4-P_{,\tau}$.  If $P_{,\tau}$ is negative after a slide, then
there is a bounce as just described above.  If $P_{,\tau} > 2$
after this bounce, then a slide follows, and so on.
In each case, finally, $0 < P_{,\tau} < 2$, $P_{,\tau} \approx v$
and the conditions are satisfied such that the explicit
approximation of \cite{beverlyandme} should be valid throughout
the remaining approach to the singularity.  These results
and all other results we obtained from various choices of
initial data\footnote{This method allows study of a large
number of different spikes with varied initial conditions 
into the ``asymptotic regime'', since the code runs very quickly.}
agree with the picture we have just presented.


The rule for the final (after all the slides and bounces) value
of $P_{,\tau}$, based on the initial value of $P_{,\tau}$ is shown in
table \ref{table1}, with $n \geq 0$ an integer, and $0 < \sigma < 1$.
Both initially and finally, $v \approx P_{,\tau}$, but this relation
does not hold during bounces and slides.  The rule for the final
value of $P_{,\tau}$ based on the initial value of $P_{,\tau}$ when
$Q_{,x} \neq 0$, obtained from the explicit approximations \cite{beverlyandme}
(if $P_{,\tau} >1$, $P_{,\tau} \to 2 - P_{,\tau}$ and then if
$P_{,\tau} < 0$, $P_{,\tau} \to -P_{,\tau}$ and reiterate)
is also given in the table.\footnote{The discussion in this paper
assumes ``genericity conditions'', for example, that
$Q_{,xx}\neq 0$ at upward pointing spikes.
For discussion of persistence of high velocity spikes under
special conditions, see \cite{alanandmarsha}.}
\begin{table}
\caption{Final $P_{,\tau}$ from initial $P_{,\tau}>0$,
for integer $n \geq 0$ and $0 < \sigma < 1$.}
\begin{center}
\begin{tabular}{|l|c|c|} \hline
Initial $P_{,\tau}$ & Final $P_{,\tau}$ if $Q_{,x} = 0$ 
& Final $P_{,\tau}$ if $Q_{,x} \neq 0$ \\ \hline
$4n + \sigma $ & $\sigma$ & $\sigma$ \\ \hline
$4n + 1 + \sigma $ & $1+\sigma$ & $1-\sigma$ \\ \hline
$4n + 2 + \sigma $ & $2-\sigma$ & $\sigma$ \\ \hline
$4n + 3 + \sigma $ & $1-\sigma$ & $1-\sigma$ \\ \hline
\end{tabular}
\end{center}
\label{table1}
\end{table}

To obtain a picture of what happens at downward pointing spikes, one
could use the same combination t-code and c-code with the appropriate
initial data.  But this is not necessary, since an upward
pointing spike is mapped to a downward pointing spike with
$P_{,\tau}^{\rm down} = 1 -P_{,\tau}^{\rm up}$ by the Gowdy to Ernst
transformation \cite{alanandmarsha}.  Thus for downward pointing
spikes an {\it upward} slide in $P_{,\tau}$ occurs if $P_{,\tau} < -1$
at the spike, and the rule for the slide is
$P_{,\tau} \to -P_{,\tau} -2$.  If $P_{,\tau} > 2$ after the slide then
there is a {\it downward} bounce $P_{,\tau} \to 2 - P_{,\tau}$.
Reiterating this until $-1 < P_{,\tau} < 1$,
we obtain the rule shown in table \ref{table2},
with $n \geq 0$ and $0 < \sigma < 1$.  The Gowdy to Ernst
transformation maps the condition $Q_{,x} = 0$ for an upward
pointing spike to form to the condition $e^{2P} Q_{,\tau} = 0$ for
a downward pointing spike to form.
\begin{table}
\caption{Final $P_{,\tau}$ from initial $P_{,\tau}<0$,
for integer $n \geq 0$ and $0 < \sigma < 1$.}
\begin{center}
\begin{tabular}{|l|c|c|} \hline
Initial $P_{,\tau}$ & Final $P_{,\tau}$ if $Q_{,\tau} = 0$ 
& Final $P_{,\tau}$ if $Q_{,\tau} \neq 0$ \\ \hline
$-4n - \sigma $ & -$\sigma$ & $\sigma$ \\ \hline
$-4n - 1 - \sigma $ & $-1+\sigma$ & $1-\sigma$ \\ \hline
$-4n - 2 - \sigma $ & $\sigma$ & $\sigma$ \\ \hline
$-4n - 3 - \sigma $ & $1-\sigma$ & $1-\sigma$ \\ \hline
\end{tabular}
\end{center}
\label{table2}
\end{table}

\begin{figure}
\includegraphics{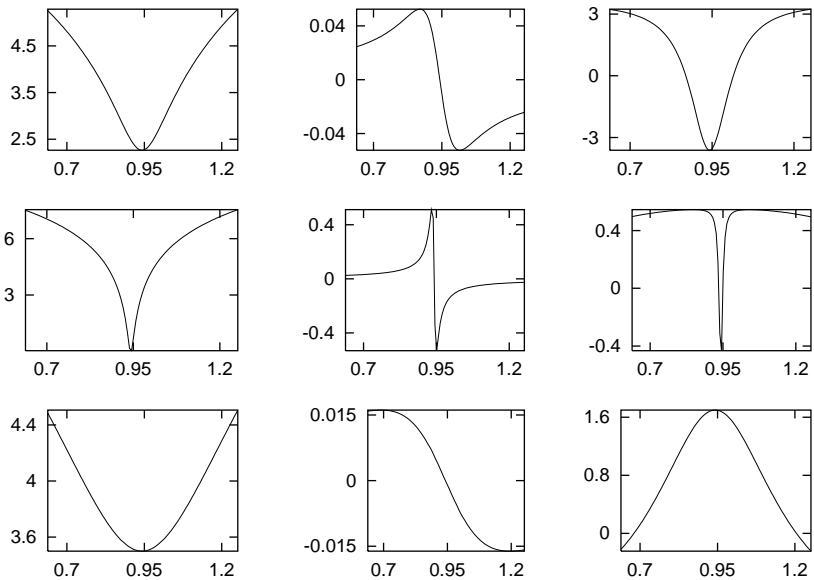}
\caption{\label{figwhole}$P$ {\it vs} $x$ (left column),
$Q$ {\it vs} $x$ (middle column) and $P_{,\tau}$ {\it vs} $x$
(right column) at $\tau \approx 3.1$ (top row),
$\tau \approx 4.0$ (middle row) and $\tau \approx 6.4$ (bottom
row).}
\end{figure}

The c-code only gives us information concerning an extremely small
neighborhood of the spike.  This is precisely its advantage.
But the rules just obtained raise the question, ``How does the
small neighborhood of the spike match onto the rest of space
when the rule for the final value of $P_{,\tau}$ at the spike agrees
with the rule for the final value of $P_{,\tau}$ elsewhere ({\it i.e.},
in half the cases)?''  To see how a spike sits in the rest of
the spacetime, we use Cauchy evolution using the symplectic PDE
solver \cite{beverlyandme} with mesh refinement \cite{marsha}.
We use a resolution of 1024 (the 1025th grid point is identified
with the 1st, corresponding to a difference in $x$ of $2 \pi$).
Two levels of mesh refinement are made at spikes, each
increasing the resolution five-fold.  This code is not as fast
as the combination t-code and c-code, and is not as effective
over as large of a range of initial data sets nor as far into
the asymptotic regime as the combination t-code and c-code.
But all the results obtained suggest that when the rule obtained for
final value of $P_{,\tau}$ at the spike matches the rule obtained
elsewhere, the signatures of the spike are effectively washed out,
for both upward and downward pointing spikes.
We show an example, for a downward pointing spike,
in figure~\ref{figwhole}.  The initial data is $n=162$ and
$m=2048$ (so $t_0 \approx 0.99$), $s = 5.7$, $p = 0$,
$q = 0.01$ and now $Q = q \cos(x + \pi/5)$ so that
no spike is at the edge of the grid, for easier implementation
of the mesh refinement.  $P_{,\tau} \approx -3.7$ when the downward
pointing spike begins to form.  The spiky features are pronounced
in the middle row of figure~\ref{figwhole}, but by the time of the
final row they are disappearing. 

\section{Conclusions}

We have seen that there is a mechanism that, in general, eventually
drives upward pointing spike velocities below 2 and downward pointing
spike velocities below 1.  The general asymptotic behavior of
Gowdy spacetimes is then that given in \cite{beverlyandme}.  The 
spacetime is everywhere AVTD.  At a set of isolated points there
are spikes.  The closer a point is to a spike,
the longer it takes to reach the VTD regime.  Upward
pointing spikes in general have $1< v < 2 $, downward pointing spikes
in general have $0< v < 1$ ($-1 < P_{,\tau} < 0$) and the shape of both kinds
is described by expressions in \cite{beverlyandme}.

\section{Acknowledgments}

This work was partially supported by NSF grant
PHY-9988790 to Oakland University and by the Natural Sciences and
Engineering Research Council of Canada.

\end{document}